\documentclass[a4paper]{jpconf}
\usepackage{graphicx}
\begin{document}
\title{The Atmospheric Monitoring Strategy for the Cherenkov Telescope Array}

\author{MK Daniel$^{1}$ 
for the CTA Consortium$^2$}

\address{
$^{1}$ Department of Physics, University of Liverpool, Liverpool, L69 7ZE. UK. \newline
$^2$ http://www.cta-observatory.org/

}

\ead{michael.daniel@liverpool.ac.uk}

\begin{abstract}
The Imaging Atmospheric Cherenkov Technique (IACT) is unusual in astronomy as 
the atmosphere actually forms an intrinsic part of the detector system, 
with telescopes indirectly detecting very high energy particles 
by the generation and transport of Cherenkov photons deep within the 
atmosphere. This means that accurate measurement, characterisation and 
monitoring of the atmosphere is at the very heart of successfully 
operating an IACT system. The Cherenkov Telescope Array (CTA) will be 
the next generation IACT observatory with an ambitious aim to improve 
the sensitivity of an order of magnitude over current facilities, 
along with corresponding improvements in angular and energy resolution 
and extended energy coverage, through an array of Large (23\,m), Medium 
(12\,m) and Small (4\,m) sized telescopes spread over an area of order 
$\sim$km$^2$. Whole sky coverage will be achieved by operating at two sites: one in the northern hemisphere and one in the southern hemisphere. This proceedings 
will cover the characterisation of the candidate sites and the 
atmospheric calibration strategy. CTA will utilise a suite of 
instrumentation and analysis techniques for atmospheric modelling 
and monitoring regarding pointing forecasts, intelligent pointing 
selection for the observatory operations and for offline data correction.
\end{abstract}

\section{Introduction}
\begin{figure}[htbp]
\begin{center}
\includegraphics[width=0.75\textwidth, trim=0cm 1cm 0cm 1cm]{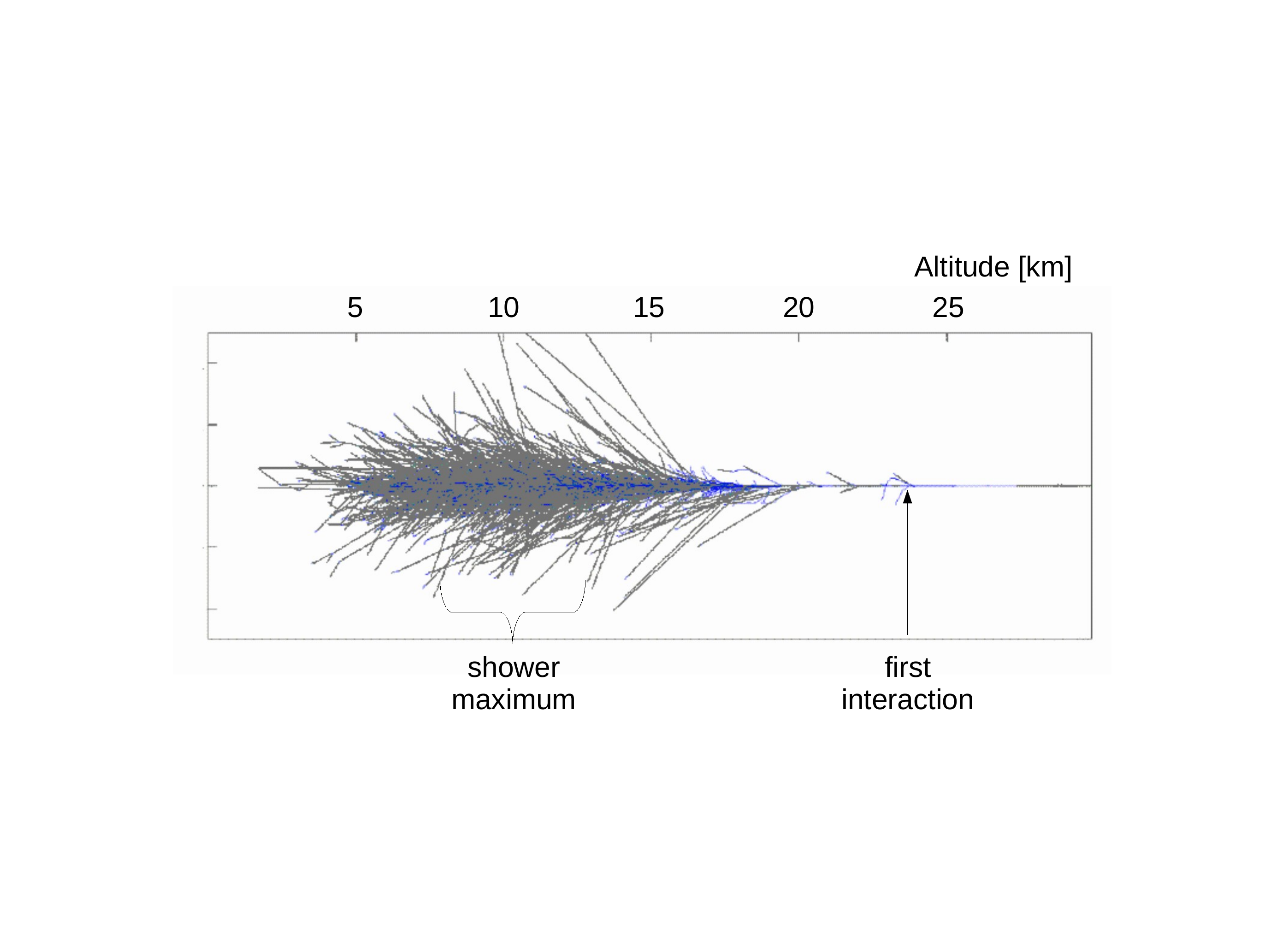}
\includegraphics[width=0.75\textwidth]{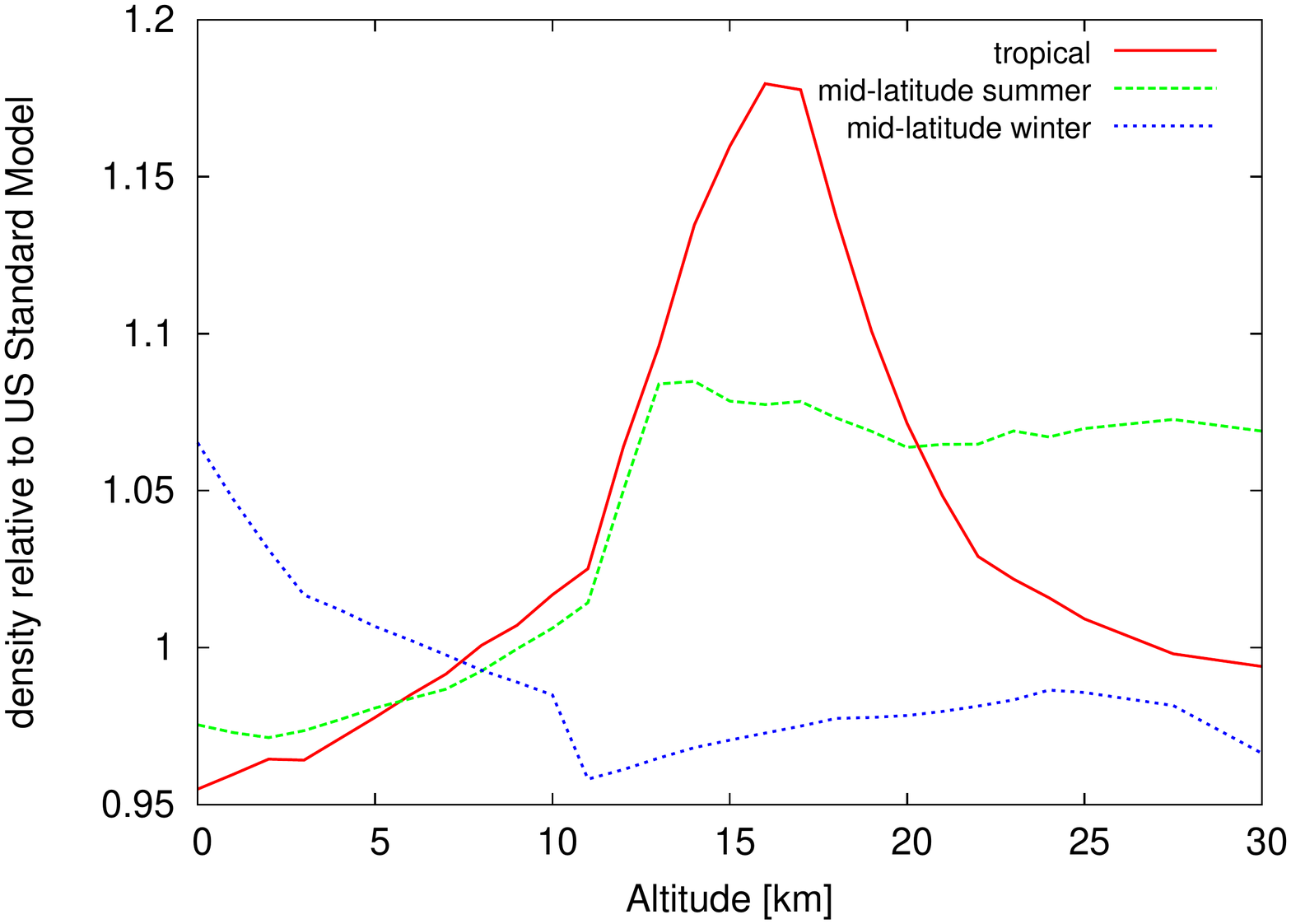}
\end{center}
\caption{\label{fig:EAS}
\textbf{Top:} A representation of an Extensive Air Shower initiated by a 1 TeV gamma-ray photon. Most of the Cherenkov light is generated below 15\,km altitude, with the shower maximum at 10\,km.
\textbf{Bottom:} The density profile of the tropical (solid red line), mid-latitude summer (dashed green line) and mid-latitude winter (dotted blue line) relative to the US Standard model showing the variation due to location and season. The variation is greatest in the development region between the first interaction and where shower maximum occurs.
}
\end{figure}

Ground based gamma-ray astronomy is not able to directly measure the 
high energy photons, for the atmosphere is thankfully\footnote{for the general purpose of daily life} opaque to them. Instead this branch of astronomy exploits an indirect method, by measuring optical photons of Cherenkov light (concentrated toward the blue end of the spectrum) produced from the faster than light-in-air passage of the particles produced in an extensive air shower (EAS) that results from the multiple interactions that redistribute the energy from a single very high energy primary particle to many lower energy ones, see figure
~\ref{fig:EAS} top. The Cherenkov light is generated high in the atmosphere (typically the first interaction is at 20\,km altitude and the maximum of emission between 8-12\,km depending on the primary energy and the atmospheric density profile) and spreads out at a characteristic angle that depends on the density of the medium (for air this is about a degree) (see fig~\ref{fig:EAS} bottom for how this varies depending on latitude and season). When Cherenkov photons that are not attenuated reach ground level they are thus fairly uniformly spread in a pool of order 100\,m radius and drop off rapidly after that. Placing a telescope anywhere within that pool of Cherenkov light means the EAS can be detected and so this provides a very large effective area for the detection of very low flux sources (a few photons per minute is a bright source). The extent of the Cherenkov image means it is immune to many of the atmosphere's characteristics that affect traditional ground based optical astronomy, such as seeing/scintillation effects, but with the atmosphere comprising such a fundamental component of the detection system, and acting as a calorimeter, it is still very much subject to changes that are beyond our control and must be monitored for and measured instead.

The Imaging Atmospheric Cherenkov Technique (IACT) is the most sensitive 
method of detecting astrophysical sources of very high energy (VHE, E$>10$ GeV) 
gamma rays. The Cherenkov Telescope Array (CTA)\footnote{www.cta-observatory.org} \cite{CTA} will be the next generation IACT facility, providing the community with an open-access observatory for the observation of gamma rays with energies from a few tens of GeV to hundreds of TeV with unprecedented sensitivity and angular and energy resolution. Whole sky coverage will be achieved by operating at sites in both hemispheres: the southern array, covering an area of about 10\,km$^{2}$ with O(100) telescopes of all sizes, is optimised for the highest energy observations of the galactic plane; and the northern array, covering an area of about 1\,km$^{2}$, is optimised for the low 
energy thresholds best suited to observing extragalactic sources.

The CTA has set some very stringent requirements, including: the uncertainty on the effective collection area to be $<12\%$; the systematic error on the absolute intensity of Cherenkov photons at each telescope (post calibration) must be $<8$\% such that the systematic error on the energy scale must not exceed 10\%; a final systematic error on the localisation of a point-like source of 3 arcseconds (under favourable observing conditions). Furthermore, the 
observatory will need to maximise the duty cycle and keep the maintenance budget low. As the atmosphere is such an integral part of the detection system for IACT, in order to achieve these ambitions the CTA requires a suitably ambitious atmospheric monitoring strategy compared to what has come before, which will be detailed in the following sections.

\section{Atmospheric Monitoring Strategy}
The goal of the CTA atmospheric monitoring strategy is to both characterise the systematic uncertainties in gamma-ray event reconstruction and to minimise any loss of data due to non-optimal atmospheric conditions. 
With an appropriate atmospheric monitoring strategy, not only can the energy and flux of an astrophysical source be correctly retrieved but also the overall duty cycle of the observatory can be increased. The CTA strategy is split into 5 tasks
\begin{description}
 \item[Climatology] Characterisation of the site conditions and simulation input. This task will begin as soon as the site decision is made, even before telescopes are installed. The use of weather stations and radiosondes will build up the database of atmospheric conditions that can be input to and tested against data models like GDAS\footnote{http://www.ncdc.noaa.gov/data-access/model-data/model-datasets/global-data-assimilation-system-gdas}, WRF\footnote{http://www.wrf-model.org/index.php} and SENES\footnote{http://www.senes.ca/} to ensure they are reliable enough descriptions of the site conditions, such as has been successfully achieved for the Auger site \cite{Will}. Also a profound knowledge of aerosol size distributions, their occurence and variability shall be obtained.
 \item[Weather Nowcast, Forecast, Alerts \& Protection] Protect the instrumentation, inform observing schedule. Instruments providing readily available information of the weather and short term forecasts are not only useful for data-taking, but for planning maintenace trips, etc to the extended array. Many of these will be commercial instruments to provide standardised reliable 24/7 remote operations and automated alerts.
 \item[Online Smart Scheduling] Find the optimal target in non-ideal observing conditions. 
 \item[Offline Data Selection] Certify good quality data. If the conditions do not allow for unaffected/unbiased observations then the recoverable data must be flagged against those which must be rejected.
 \item[Offline Data Correction] Retrieve correct flux and energy estimates. Data taken during some non-optimal atmospheric conditions can be corrected by the use of specially adapted Monte Carlo simulations or clever algorithms making use of the atmospheric data. 
\end{description}
Equipment is primarily designed to serve one of the tasks, but may be used for others as well, e.g. for cross-checks and calibration. Where possible passive monitoring devices that do not interfere with Cherenkov observations or take away from source observation time are preferred, but not always possible.

\section{Atmospheric Monitoring Equipment}
A mixture of custom built and commercial instruments will comprise the bulk of the atmospheric monitoring equipment.

\subsection{ATMOSCOPE}
The ATMOSCOPE (Autonomous Tool for Measuring Site COndition PrEcisely) \cite{ATMOSCOPE} is a sensor station comprised of 
\begin{itemize}
 \item a commercial weather station for measuring temperature, humidity, pressure and wind;
 \item a Light of Night Sky (LoNS) instrument for measuring the sky brightness through B and V band filters;
 \item an All Sky Camera for monitoring night sky quality.
\end{itemize}
designed to be deployed and operated in a remote site without power supply and ethernet connection. The ATMOSCOPEs were cross-calibrated and then deployed to each of the CTA site candidates to provide an independent and unbiased comparison for the purpose of site selection. In addition the units can be compared to existing devices at sites and satellite data to provide a better understanding of long term historical datasets.

\subsection{All Sky Camera}
The All Sky Camera (ASC) is an astronomical CCD camera with a fish-eye lens to determine the night sky quality from the cloud fraction (percentage of the sky covered by clouds) and the night sky brightness (in mag/arcsec$^{2}$) \cite{ASC}. An upgraded system to the one used in the site selection study will be used to estimate sky quality and determine areas of the sky free from cloud for the CTA central scheduler.

\subsection{FRAM}
The F/(Ph)otometric Robotic Atmospheric Monitor (FRAM) \cite{FRAM} is a wide field astronomical imaging camera on a robotic mount to monitor stars in the whole field of view of the Cherenkov telescopes (up to $10^\circ \times 10^\circ$). From measuring the light flux of the stars in the same direction we can determine the atmospheric extinction with high temporal and spatial resolution without interfering with the Cherenkov observations.

\subsection{UVscope}
The UVscope \cite{UVscope} is a portable, NIST-calibrated, multi-pixel photon detector developed at INAF IASF-Palermo (Italy) to support experimental activities in the fields of high-energy astrophysics and cosmic-rays, but thanks to its features and operational flexibility the instrument can be used in a wide range of applications. The instrument works in single photon counting mode to directly measure the light flux in the 300-650\,nm wavelength range. The atmospheric transparency can be evaluated by tracking a star measuring the absorption of the flux as a function of the air mass travelled. The UVscope can also measure the diffuse emission of the night-sky-background (NSB) in the same field of view of a given telescope \cite{UVscopeAuger}. In the CTA framework, UVscope would follow the same pointing of a given CTA telescope or group of telescopes with a quantum efficiency well matched in the wavelength region of interest for Cherenkov light detection \cite{ASTRI}.

\subsection{Cherenkov Transparency Coefficient}
A Cherenkov telescope is always going to be the most sensitive instrument to the weather conditions that affect Cherenkov light yield. The Cherenkov Transparency Coefficient (CTC) method \cite{CTC} combines the information of the average pixel gain and optical throughput of the telescope to the expected telescope trigger rates for cosmic-ray protons to determine if the trigger energy threshold of the telescope has changed due to atmospheric effects. This enables the CTC to be used as a direct and free data quality selection parameter, but will naturally suffer from the fact it can only do so post data acquisition. A high correlation factor of the CTC with independent aerosol optical depth measurements by the MISR satellite and onsite lidar and IR radiometer systems at the H.E.S.S. site in Namibia means such independent systems could be used to automatically interact with the central autoscheduler to determine an optimum pointing position. For full offline data correction, though, a system providing height information is required.

\subsection{Lidar}
Because the Cherenkov light is generated and transported within the atmosphere just treating it like a simple filter can lead to a biased estimate of the amount of attenuation \cite{konrad}. Instead a differential transmission profile is needed, meaning instrumentation with height resolution, for which a lidar is best suited. Also, something capable of distinguishing between different types of attenuation, from the Rayleigh scattering of the molecular component of the atmosphere and the Mie scattering of aerosols, is highly desirable for CTA to reach its goal energy resolution and energy scale bias requirements. A Raman lidar should enable us to decouple the molecular and aerosol components to $\sim5\%$ uncertainty. 
Several institutes in CTA are involved with developing Raman lidar systems. Generically these lidars are characterised by having large reflectors of $\geq1.5$\,m diameter, powerful Nd:YAG lasers operated at 2-3 wavelengths (1054\,nm, 532\,nm, 355\,nm) with filters at 355\,nm and 532\,nm for the elastic backscatter component and 387\,nm and 607\,nm for the N$_{2}$ inelastic scatter Raman lines. The different groups are prototyping independent mechanical, optical and steering solutions. The Laboratoire Univers et Particules de Montpellier, Institut de Fisica d'Altes Energies (IFAE) and the Universitat Aut\`{o}noma de Barcelona (UAB) are re-purposing some CLUE Cherenkov telescopes for this purpose; the Centro de Investiciones Laser y sus Aplicaciones (CEILAP) group have a custom designed solution based on a multiple mirror system. More details of these devices can be found in \cite{lidars}. Lidar laser shots will affect the acquisition of Cherenkov events, but will occur at low frequency ($\sim20$\,Hz) and can be easily flagged and removed in the post-observation data analysis.

\subsubsection{Ceilometer}
Primarily used by the aviation industry and designed to be eye safe and remotely operated, a ceilometer is a low power version of a lidar utilising an infra-red (905\,nm) laser diode and summing many thousand return traces to detect the height of clouds. A Vaisala CL51 model ceilometer has the capability to report up to three cloud layers in a range up to 13\,km and gives backscatter profiling for vertical visibility up to 15\,km. In favourable conditions the backscatter profile can also be used to monitor boundary layer structures \cite{ceil}. Use of an infra-red device means that the heights of clouds anywhere in the sky can be determined without interference to Cherenkov observations. If the height of the cloud is above shower maximum then data can still be collected with minimal bias, for lower cloud then appropriate data recovery techniques could be applied~\cite{pks2155, garrido2013}.

\subsection{Global Networks, Satellite Data and Atmospheric Models}
Satellites such as MODIS, GOES and METEOSAT are available to offer accurate measurements in the regions not (yet) covered by ground-based instrumentation and can fill in the long term climatology history for a region, albeit at coarse scale and, unless geostationary, often low cadence. Ground-based atmospheric monitoring networks provide input to global atmospheric models such as GDAS, WRF or SENES. Networks are also important because they impose standardised algorithms, thus reducing the potential for bias. AERONET is a network of Cimel sun photometers installed at some 550 stations worldwide that provides the most accurate measurement of aerosol optical depth, to $\delta \tau \sim 1\%$.
CTA is investigating the mutual benefits both of profiting from the wealth of data provided by these networks and from providing access to the data of the atmospheric monitoring equipment of the CTA observatory, given these will be located at remote sites normally not covered by ground-based installations.

\section{Summary}
The requirements for the next generation ground based gamma-ray observatory to have unprecedented sensitivity in flux estimation and minimal uncertainties in energy reconstruction also demands a comprehensive atmospheric monitoring strategy. The suite of instruments planned for the CTA sites will perform pointed observations and provide all-sky characterisations that will provide an accurate site climatology, enable smart online scheduling in addition to offline data selection and correction where necessary. This will serve to not only minimise systematic errors, but also boost the useful operational duty cycle of CTA.

\ack
We gratefully acknowledge support from the agencies and organizations 
listed under Funding Agencies at this website: http://www.cta-observatory.org/.

\section*{References}
\bibliographystyle{iopart-num}
\bibliography{cta-ams}

\providecommand{\newblock}{}
\begin{thebibliography}{10}
\expandafter\ifx\csname url\endcsname\relax
  \def\url#1{{\tt #1}}\fi
\expandafter\ifx\csname urlprefix\endcsname\relax\def\urlprefix{URL }\fi
\providecommand{\eprint}[2][]{\url{#2}}

\bibitem{CTA}
{Actis} M, {Agnetta} G, {Aharonian} F, {Akhperjanian} A, {Aleksi{\'c}} J,
  {Aliu} E, {Allan} D, {Allekotte} I, {Antico} F, {Antonelli} L~A, {Antoranz}
  P, {Aravantinos} A, {Arlen} T, {Arnaldi} H, {Artmann} S, {Asano} K, {Asorey}
  H, {B{\"a}hr} J, {Bais} A, {Baixeras} C and et~al 2011 {\em Experimental
  Astronomy\/} {\bf 32} 193--316 (\textit{Preprint} \eprint{1008.3703})

\bibitem{Will}
{Keilhauer} B and {Will} M 2012 {\em European Physical Journal Plus\/} {\bf
  127} 96 (\textit{Preprint} \eprint{1208.5417})

\bibitem{ATMOSCOPE}
{Vincent} S and {CTA Consortium} f~t 2014 {\em ArXiv e-prints\/}
  (\textit{Preprint} \eprint{1403.5075})

\bibitem{ASC}
{Mandat} D, {Pech} M, {Hrabovsky} M, {Schovanek} P, {Palatka} M, {Travnicek} P,
  {Prouza} M and {Ebr} J 2014 {\em ArXiv e-prints\/} (\textit{Preprint}
  \eprint{1402.4762})

\bibitem{FRAM}
{Prouza} M, {Jel{\'{\i}}nek} M, {Kub{\'a}nek} P, {Ebr} J, {Tr{\'a}vn{\'{\i}}{\v
  c}ek} P and {{\v S}m{\'{\i}}da} R 2010 {\em Advances in Astronomy\/} {\bf
  2010} 849382

\bibitem{UVscope}
{Maccarone} M~C, {Catalano} O, {Giarrusso} S, {La Rosa} G, {Segreto} A,
  {Agnetta} G, {Biondo} B, {Mangano} A, {Russo} F and {Billotta} S 2011 {\em
  Nuclear Instruments and Methods in Physics Research A\/} {\bf 659} 569--578

\bibitem{UVscopeAuger}
{Segreto} A 2011 {\em International Cosmic Ray Conference\/} {\bf 3} 129

\bibitem{ASTRI}
{Maccarone} M~C, {Segreto} A, {Catalano} O, {La Rosa} G, {Russo} F, {Sottile}
  G, {Gargano} C, {Biondo} B, {Fiorini} M, {Incorvaia} S and {Toso} G 2014 {\em
  Society of Photo-Optical Instrumentation Engineers (SPIE) Conference
  Series\/} ({\em Society of Photo-Optical Instrumentation Engineers (SPIE)
  Conference Series\/} vol 9149)

\bibitem{CTC}
{Hahn} J, {de los Reyes} R, {Bernl{\"o}hr} K, {Kr{\"u}ger} P, {Lo} Y~T~E,
  {Chadwick} P~M, {Daniel} M~K, {Deil} C, {Gast} H, {Kosack} K and {Marandon} V
  2014 {\em Astroparticle Physics\/} {\bf 54} 25--32 (\textit{Preprint}
  \eprint{1310.1639})

\bibitem{konrad}
{Bernl{\"o}hr} K 2000 {\em Astroparticle Physics\/} {\bf 12} 255--268
  (\textit{Preprint} \eprint{astro-ph/9908093})

\bibitem{lidars}
{Doro} M, {Gaug} M, {Pallotta} J, {Vasileiadis} G, {Blanch} O, {Chouza} F,
  {D'Elia} R, {Etchegoyen} A, {Font} L, {Garrido} D, {Gonzales} F,
  {L{\'o}pez-Oramas} A, {Mart{\'{\i}}nez} M, {Otero} L, {Quel} E, {Ristori} P
  and {for the CTA consortium} 2014 {\em ArXiv e-prints\/} (\textit{Preprint}
  \eprint{1402.0638})

\bibitem{ceil}
{M{\"u}nkel} C 2006 {\em Society of Photo-Optical Instrumentation Engineers
  (SPIE) Conference Series\/} ({\em Society of Photo-Optical Instrumentation
  Engineers (SPIE) Conference Series\/} vol 6367)

\bibitem{pks2155}
{Nolan} S~J, {P{\"u}hlhofer} G and {Rulten} C~B 2010 {\em Astroparticle
  Physics\/} {\bf 34} 304--313 (\textit{Preprint} \eprint{1009.0517})

\bibitem{garrido2013}
{Garrido} D, {Gaug} M, {Doro} M, {Font} L, {L\'opez-Oramas} A and {Moralejo} A
  2013 {\em {Proceedings of the 33$^\mathrm{rd}$ ICRC, Rio de Janeiro}\/}
  {arXiv:1308.0473}

\end{thebibliography}

\end{document}